\documentclass[twocolumn, 
                          showpacs, 
                          nofootinbib, 
                          nobibnotes, 
                          aps,  
                          superscriptaddress, 
                          amssymb, 
                          amsmath, 
                          floatfix, 
                          reprint, 
                          pra, 
                          notitlepage, 
                          showkeys,
                          10pt]{revtex4-2}

\usepackage{blindtext}
\usepackage{amssymb}
\usepackage{amsmath}
\usepackage{graphicx}
\usepackage{dcolumn}
\usepackage{url}
\usepackage[colorlinks=true,breaklinks=true,allcolors=blue]{hyperref}
\usepackage{tikz}
\usepackage{dsfont}
\usepackage{epsfig}
\usepackage{color}
\usepackage[ruled]{algorithm2e}
\usepackage{bbold}
\usepackage[binary-units=true]{siunitx}
\usepackage[acronym]{glossaries}
\usepackage{tikz}
\usetikzlibrary{calc}
\usepackage{bm}
\usepackage{cleveref}
\usepackage{placeins}
\usepackage{changes}

\usetikzlibrary{decorations.text}

\usetikzlibrary{shapes.symbols, shapes.arrows, calc, shapes}

\def\equationautorefname~#1\null{Equation~(#1)\null}

\glsdisablehyper
\newacronym{MCMC}{MCMC}{Markov-chain Monte Carlo}
\newacronym{TFIM}{TFIM}{transverse field Ising model}
\newacronym{RBM}{RBM}{restricted Boltzmann machines}
\newacronym{BSS2}{BSS-2}{BrainScaleS-2}
\newacronym{LIF}{LIF}{leaky integrate-and fire}
\newacronym{SNN}{SNN}{spiking neural network}
\newacronym{DKL}{$D_\mathrm{KL}$}{Kullback-Leibler divergence}
\newacronym{adex}{AdEx}{adaptive exponential leaky integrate-and-fire}

\renewcommand{\vec}[1]{\bm{#1}}

\newcommand{\el}{V_\mathrm{l}}
\newcommand{\gl}{g_\mathrm{l}}
\newcommand{\cm}{C_\mathrm{m}}
\newcommand{\isyn}{I_\mathrm{syn}}
\newcommand{\vthresh}{V_\mathrm{thresh}}
\newcommand{\vreset}{V_\mathrm{reset}}

\newcommand{\DKL}{D_\mathrm{KL}}

\newcommand{\ket}[1]{|#1\rangle}
\newcommand{\bra}[1]{\langle#1|}

\newcommand{\tauref}{\tau_{\mathrm{ref}}}

\crefname{equation}{Eqn.}{Eqns.}
\Crefname{equation}{Eqn.}{Eqns.}
\crefname{figure}{Fig.}{Figs.}
\Crefname{figure}{Fig.}{Figs.}
\crefname{section}{Sec.}{Secs.}
\Crefname{section}{Sec.}{Secs.}

\crefformat{app}{methods #2#1#3}
\crefrangeformat{app}{methods #3#1#4 to #5#2#6}
\crefmultiformat{app}{methods #2#1#3}{ and #2#1#3}{, #2#1#3}{ and #2#1#3}
\crefrangemultiformat{app}{methods #3#1#4 to #5#2#6}{ and #3#1#4 to #5#2#6}{, #3#1#4 to #5#2#6}{ and #3#1#4 to #5#2#6}
\Crefformat{app}{Methods #2#1#3}
\Crefrangeformat{app}{Methods #3#1#4 to #5#2#6}
\Crefmultiformat{app}{Methods #2#1#3}{ and #2#1#3}{, #2#1#3}{ and #2#1#3}
\Crefrangemultiformat{app}{Methods #3#1#4 to #5#2#6}{ and #3#1#4 to #5#2#6}{, #3#1#4 to #5#2#6}{ and #3#1#4 to #5#2#6}

\makeatletter
\let\cat@comma@active\@empty
\makeatother

\begin{document}

\title{Variational learning of quantum ground states on spiking neuromorphic hardware}

\author{Robert Klassert}
\affiliation{Kirchhoff-Institut f\"ur Physik, Ruprecht-Karls-Universit\"at Heidelberg, Im Neuenheimer Feld 227, 69120 Heidelberg, Germany}

\author{Andreas Baumbach}
\affiliation{Kirchhoff-Institut f\"ur Physik, Ruprecht-Karls-Universit\"at Heidelberg, Im Neuenheimer Feld 227, 69120 Heidelberg, Germany}
\affiliation{Department of Physiology, University of Bern, 3012 Bern, Switzerland}

\author{Mihai A. Petrovici}
\affiliation{Department of Physiology, University of Bern, 3012 Bern, Switzerland}
\affiliation{Kirchhoff-Institut f\"ur Physik, Ruprecht-Karls-Universit\"at Heidelberg, Im Neuenheimer Feld 227, 69120 Heidelberg, Germany}

\author{Martin G\"arttner}
\affiliation{Physikalisches Institut, Universit\"at Heidelberg, Im Neuenheimer Feld 226, 69120 Heidelberg, Germany}
\affiliation{Kirchhoff-Institut f\"ur Physik, Ruprecht-Karls-Universit\"at Heidelberg, Im Neuenheimer Feld 227, 69120 Heidelberg, Germany}
\affiliation{Institut f\"ur Theoretische Physik, Ruprecht-Karls-Universit\"at Heidelberg, Philosophenweg 16, 69120 Heidelberg, Germany}

\date{\today}

\begin{abstract}

Recent research has demonstrated the usefulness of neural networks as variational ansatz functions for quantum many-body states.
However, high-dimensional sampling spaces and transient autocorrelations confront these approaches with a challenging computational bottleneck.
Compared to conventional neural networks, physical-model devices offer a fast, efficient and inherently parallel substrate capable of related forms of Markov chain Monte Carlo sampling. 
Here, we demonstrate the ability of a neuromorphic chip to represent the ground states of quantum spin models by variational energy minimization. 
We develop a training algorithm and apply it to the transverse field Ising model, showing good performance at moderate system sizes ($N\leq 10$). 
A systematic hyperparameter study shows that scalability to larger system sizes mainly depends on sample quality, which is limited by temporal parameter variations on the analog neuromorphic chip.
Our work thus provides an important step towards harnessing the capabilities of neuromorphic hardware for tackling the curse of dimensionality in quantum many-body problems.
\end{abstract}

\maketitle

\section{Introduction}

The Hilbert space of quantum many-body systems and consequently the computational resources required to describe them grow exponentially with system size.
On the one hand, this poses a challenge to understanding collective quantum effects, for example in condensed matter physics \cite{Avella2012, Zhou2021}.
On the other hand, efficient numerical tools are required for the characterization and validation of quantum devices such as digital quantum computers currently under development \cite{preskillQuantumComputingNISQ2018a}.
Fortunately, many physical systems exhibit symmetries and structure that allow to reduce the exponential complexity and to design tractable approaches for the representation of the wave function.
For example, so-called stoquastic Hamiltonians are known to have positive ground state wave functions allowing the application of quantum Monte Carlo methods \cite{Becca2017}.
Locally interacting systems featuring an excitation gap have limited ground state entanglement, which renders tensor network states an efficient method for approximating them \cite{Schollwoeck2011}.
Such physical structure may, however, not always be easy to discover and exploit.
As the process of automatically discovering structure despite the curse of dimensionality is a discipline of machine learning, variational approaches using artificial neural networks (ANNs) have found their way into quantum many-body physics in recent years \cite{Carrasquilla2020}.
These so-called neural quantum states (NQS) have been shown to serve as efficient function approximators that rival competing approaches like tensor networks by providing accurate representations for a large class of quantum states using only a small number of parameters.
Among other applications NQS have been successfully employed as variational ans\"atze for ground state search \cite{Carleo2017, Jia2019, Carrasquilla2020}, quantum dynamics \cite{Carleo2017, Czischek2018, Schmitt2020, Hartmann2019, Nagy2019, rehTimedependentVariationalPrinciple2021}, and quantum state tomography \cite{Torlai2018, Carrasquilla2019, Torlai2019a}.

The most successful existing variational approaches for representing many-body ground states rely on the use of \gls{MCMC} methods to generate samples based on which observables are estimated \cite{melkoRestrictedBoltzmannMachines2019a}.
Probabilistic inference with \gls{MCMC} in high-dimensional spaces comes with a number of associated challenges such as trading off accuracy against sample correlations and capturing multi-modality within short simulation times.
In particular, the sampling of neural network quantum states is known to be a computationally challenging task in the case of \gls{RBM} \cite{Long2010}.
To tackle this challenge we use a physical neurmorphic device which enables the fast generation of independent samples to approximate quantum wave functions.

We develop and demonstrate a method for approximating the ground states of quantum spin systems by variationally adapting the physical parameters of a neuromorphic hardware chip.
The neuromorphic chip functions as a \gls{SNN} emulator.
Such networks work in a similar way to neuronal networks in the brain.
We use the refractory state of a neuron (refractory, $z=1$, or non-refractory, $z=0$) to encode the state (up, $\uparrow$, or down, $\downarrow$) of a quantum spin.
\gls{SNN}s, in contrast to ANNs, have inherent time dynamics and process their inputs in an event-based fashion.
Due to the physical implementation the emulation becomes inherently parallel, rendering the sampling speed independent of the network size.
We note that neuromorphic hardware has recently been used to represent entangled quantum states using a mapping of general mixed quantum states to a probabilistic representation and training the system to represent a given state by approximating its corresponding probability distribution \cite{czischekSpikingNeuromorphicChip2021}.
Here, instead, we directly encode the wave function of pure quantum states and use this approach for variational ground state search through minimization of the quantum system's total energy.
Our state representation assumes positive real wave function coefficients, a property which is guaranteed for ground states of stoquastic Hamiltonians \cite{bravyiComplexityStoquasticLocal2007}.
Using the \gls{TFIM} as a benchmark case, we find that its ground state can be represented accurately for any value of the transverse field including the quantum phase transition point.

A systematic hyperparameter study shows that the limiting factors at the current stage are of technical nature and can be overcome by an improved neuromorphic backend.
Note that, unlike other functional tasks that \gls{SNN}s have been employed for in the past \cite{Petrovici2016a,Kungl2019,Dold2019}, which only require the reproduction of large scale features, for e.g. image classes, we require the full probability distribution to be sampled with high precision.
We therefore demonstrate a new level of sampling precision for neuromorphic systems, which potentially opens up new applications beyond the specific one considered here.
Our work thus serves as a demonstration of variational ground state learning on neuromorphic devices.
This opens the door to adaptions using alternative, analog or digital neuromorphic hardware \cite{Thakur2018, Roy2019, Davies2018}, and the development of improved learning algorithms exploiting fast neuromorphic sample generation.

The remainder of this work is structured as follows:
We begin by laying the foundations of spike-based computing (\Cref{sec:background_A}) and the \acrlong{BSS2} neuromorphic substrate  (\Cref{sec:background_B}), followed by details about the variational algorithm, quantum state representation (\Cref{sec:background_C}) and the physical system, namely the \gls{TFIM} (\Cref{sec:background_D}), which it is applied to. 
In \Cref{sec:results} we examine and discuss the performance of our approach and specifically investigate the dependence on system size. 
\Cref{sec:limitations} provides a detailed analysis of the impact of hardware constraints on the performance of our method.
We conclude in \Cref{sec:conclusion} and describe future research directions.

\section{Theoretical and experimental methodology}\label{sec:background}

\begin{figure}[t]
    \centering
    \begin{tikzpicture}[]
        \begin{scope}
            \node[anchor=north west] (upper) at (0.0, 0.0) {
                \includegraphics[width=0.47\textwidth]{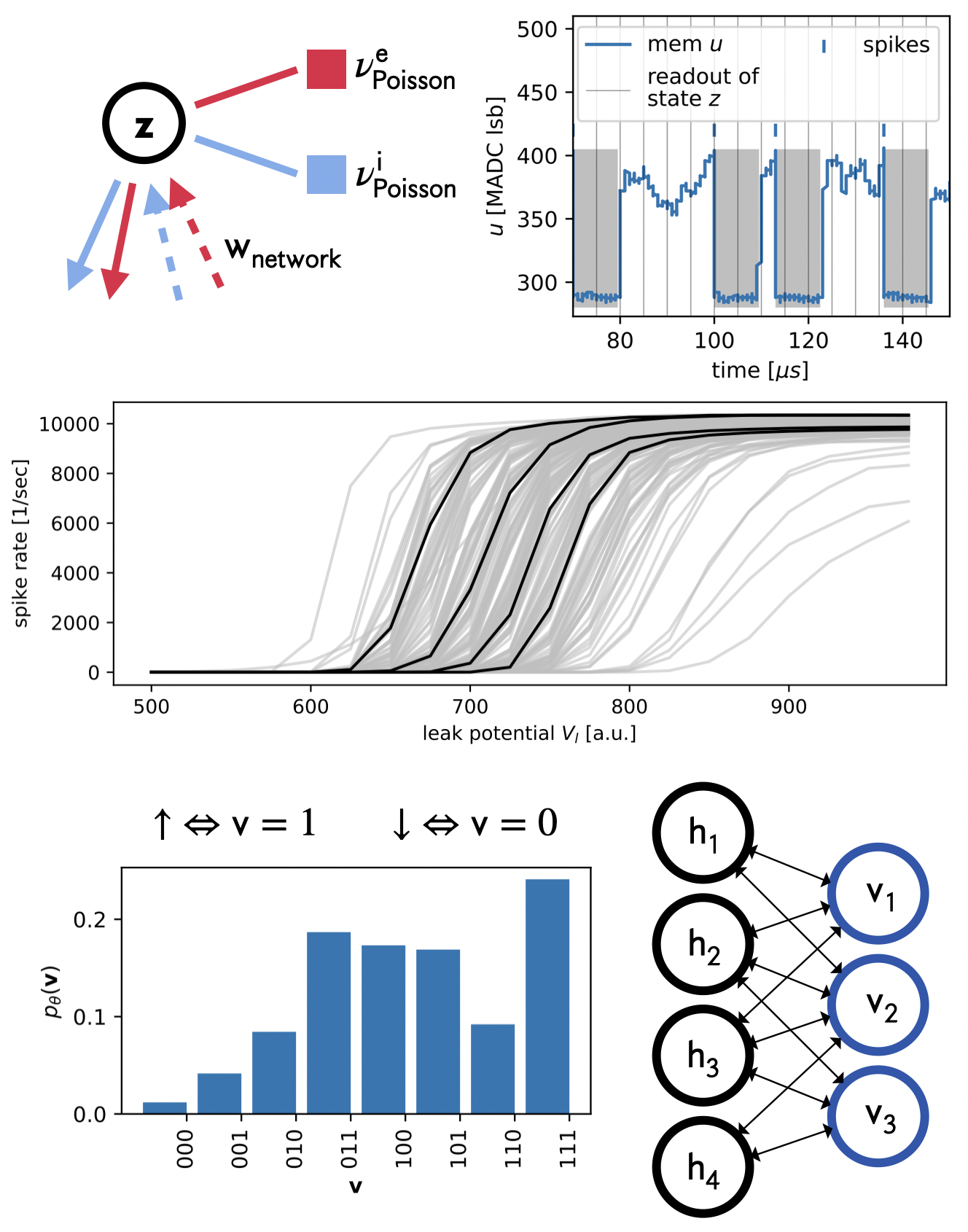}
                };

            \node[font=\bf,xshift=0.7cm,yshift=-0.1cm] at (upper.north west) {a};
            \node[font=\bf,xshift=4.5cm,yshift=-0.1cm] at (upper.north west) {b};
            \node[font=\bf,xshift=0.2cm,yshift=-3.5cm] at (upper.north west) {c};
            \node[font=\bf,xshift=0.2cm,yshift=-7.cm] at (upper.north west) {d};
            \node[font=\bf,xshift=5.5cm,yshift=-7.cm] at (upper.north west) {e};
        \end{scope}
    \end{tikzpicture}
    \caption{
        \textbf{(a)}~A LIF neuron under Poisson stimulus forms a spiking sampling unit.
        For technical reasons excitatory (red) and inhibitory (blue) connections are implemented separately.
        \textbf{(b)}~Exemplary membrane potential evolution of a spiking sampling unit.
        Binary states are assigned according to the refractory state (shaded area $z=1$, $z=0$ otherwise), which overrides the membrane dynamics after emitting a spike (blue dashes).
        States are read out periodically (grey lines).
        \textbf{(c)}~Neuronal response functions of of all 196 neurons used.
        For better visibility four of these are plotted in black.
        Note that this diversity is due to the variability of the analog substrate; for a more in-depth discussion, we refer to \cite{pfeil2013six,petrovici2014characterization,schmitt2017classification}.
        \textbf{(d)}~Frequency of occurrence of neuron states retrieved as described in panel (b) approximating the model distribution $p_\theta(\vec{v})$.
        The visible states $\vec{v}\in\{0, 1\}^N$ are identified with basis states $\ket{\vec{v}}\in\{\ket{\!\downarrow}, \ket{\!\uparrow}\}^{\otimes N}$ of the corresponding quantum spin system.
        \textbf{(e)}~Layered network architecture used throughout this manuscript.
    }
    \label{fig:1}
\end{figure}
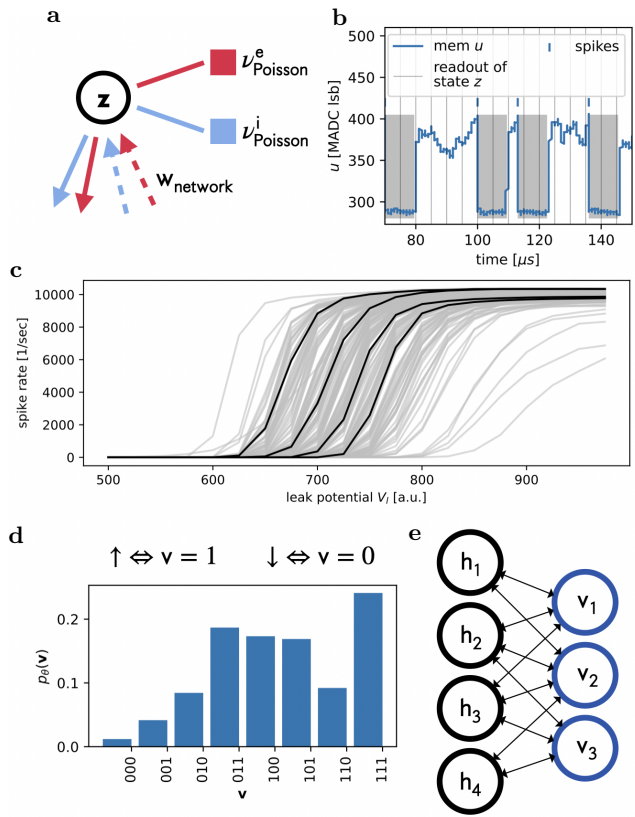

\begin{figure}[t]
    \centering
    \begin{tikzpicture}[]
        \begin{scope}
            \node[anchor=north west, xshift=0.2cm] (learning) at (0.0, 0.0) {
                \includegraphics[width=0.42\textwidth]{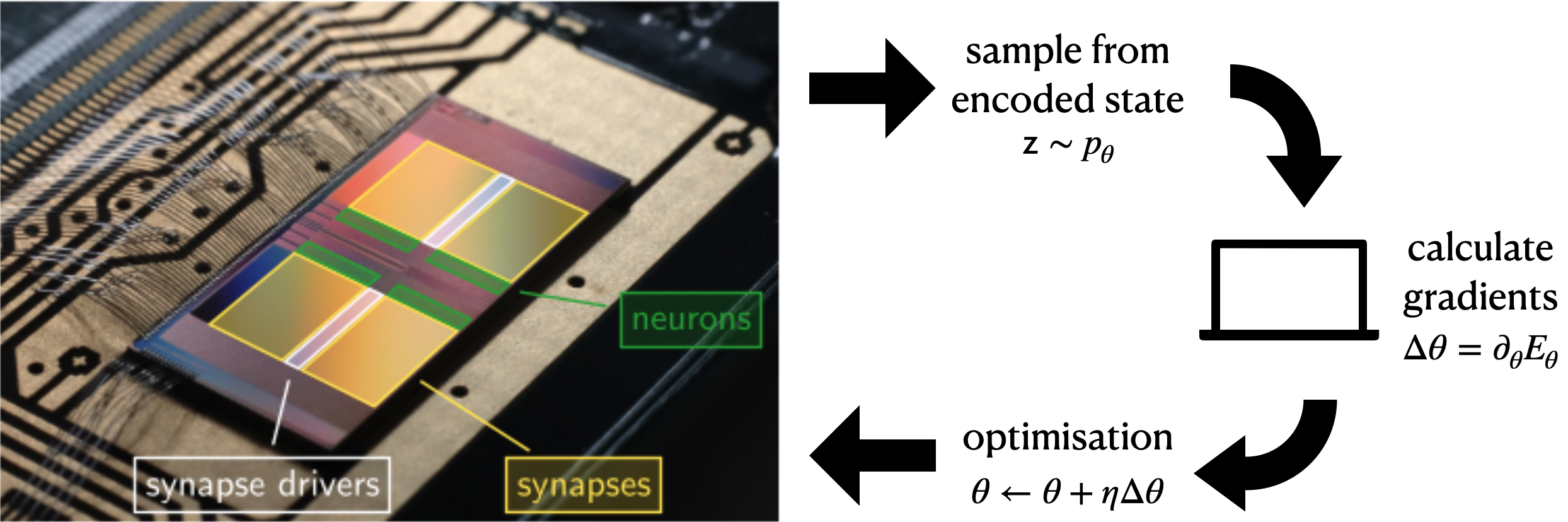}
                };
            \node[anchor=north west, xshift=-0.2cm] (figure) at (learning.south west) {
                \includegraphics[width=0.45\textwidth]{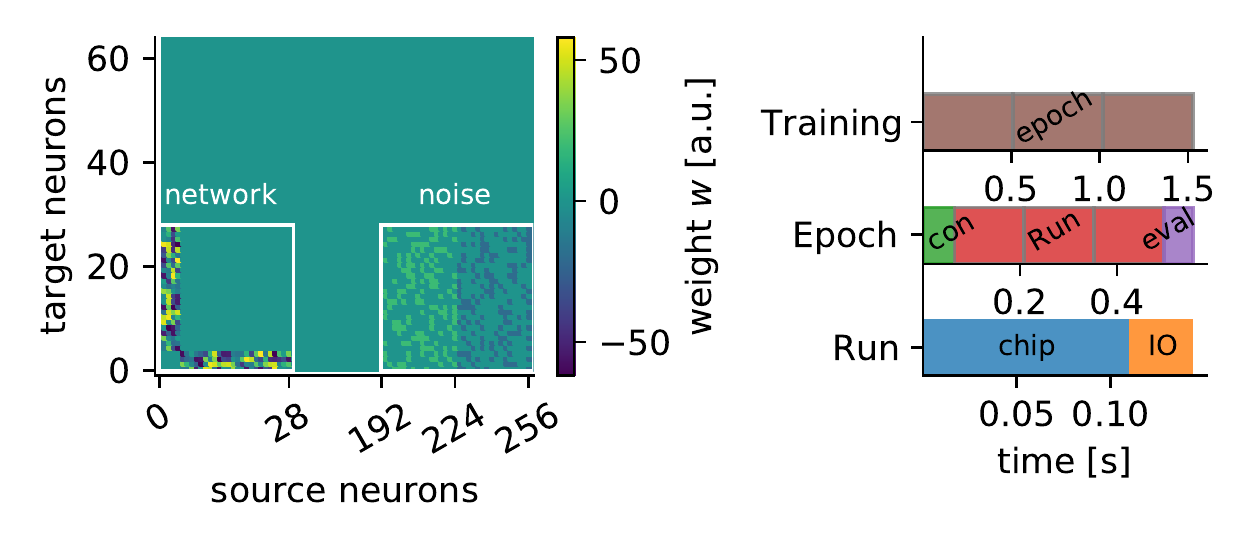}
                };

            \node[font=\bf,xshift=-0.2cm,yshift=0.cm] at (learning.north west) {a};
            \node[font=\bf,xshift=0cm,yshift=-0.2cm] at (figure.north west) {b};
            \node[font=\bf,xshift=4.9cm,yshift=-0.2cm] at (figure.north west) {c};
        \end{scope}
    \end{tikzpicture}
    \caption{
        \textbf{(a)} BrainScaleS-2 neuromorphic chip.
        It emulates the accelerated dynamics of up to 512 spiking \gls{LIF} neurons.
        The learning algorithm alternates between on-chip neural sampling and off-chip gradient calculation that informs the network parameter updates $\Delta \theta$ to minimize the energy of the represented state.
        \textbf{(b)}~Exemplary synaptic weight matrix~$\bf{w}$ for $N=8$ and $N_h=20$.
        Unused network parts (inputs 28 to 195 and neurons 64 to 195) are omitted for better visibility.
        The layered network structure manifests itself in the block structure of the lower left connectivity matrix.
        Each neuron is randomly assigned 10 out of the 64 possible noise sources (right part).
        \textbf{(c)}~Distribution of the wall clock time spent during an experimental run.
        Each epoch (brown) starts with a (partial) reconfiguration of the chip (con, green), followed by 3 consecutive sampling runs (red), followed by the evaluation (eval, purple) which includes the gradient calculation.
        Each hardware run consists of actual chip execution (chip, blue) and a transfer to the host (IO, orange).
        }
    \label{fig:2}
\end{figure}
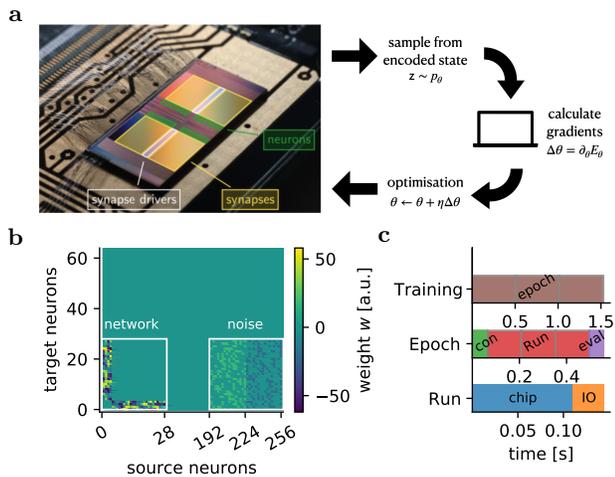

\subsection{Spike-based sampling}\label{sec:background_A}

Generative models based on artificial neural networks can be used to encode and sample from probability distributions \cite{ackleyLearningAlgorithmBoltzmann1985, Hinton1995}. 
Similarly, spiking neural networks can be shown to approximately implement Markov-chain Monte-Carlo sampling, albeit with dynamics that differ fundamentally from standard statistical methods \cite{Petrovici2016}.
Here, we use the \acrlong{BSS2} neuromorphic platform \cite{Billaudelle2019} to encode the wavefunction of quantum spin systems using a the activity distribution of a two-layer network architecture (\Cref{fig:1}).
The implementation is inspired by Boltzmann machines (BM) in that the $n$ network neurons encode binary values.
The visible units $\vec{v}=(v_1,\ldots, v_N)$ are used to directly represent the quantum spin system and the hidden units $\vec{h}=(h_1,\ldots, h_{N_h})$ mediate correlations between spins.
The full network state is the concatenation of visible and hidden units $\vec{z}=(\vec{v}, \vec{h})$.

We use \acrfull{LIF} neurons to implement our network.
The dynamics of such neurons are governed by
\begin{equation}
    \label{eq:lif}
    \cm \frac{du}{dt} = \gl \left(\el - u\right) + \isyn(t)\;,
\end{equation}
where $\cm$ is the capacitance of the neuron's membrane, $u$ its potential, $\el$ the leak potential it decays towards via the leak conductance $\gl$, and $\isyn$ the total input current to the neuron.
The input current is given by a weighted sum over spike-triggered interaction kernels $\kappa(t)$ for all spikes from all connected neurons.
For a detailed discussion we refer to \cref{sec:a_lif} and 
\cite{gerstnerSpikingNeuronModels2002}.

Whenever the membrane potential~$u$ of a neuron exceeds a threshold value~$\vthresh$, it generates a spike and the membrane is short-circuited to a reset value $\vreset < \vthresh$ (\Cref{fig:1}b).
This reset implements the fixed refractory time $\tauref$ during which we consider the neuron to be in state $z=1$ (gray shaded region in \Cref{fig:1}b, state $z=0$ otherwise).
The generated spikes are routed to other neurons via synapses with interaction strength $w$.

Networks of such \gls{LIF} neurons under Poisson stimulus (\Cref{fig:1}a) can be shown to approximately sample from characteristic Boltzmann distributions \cite{Petrovici2016}.
In this scenario, biological neurons enter a high-conductance state with a short membrane time constant $\tau_\mathrm{m}=C_m / g_l \ll \tauref$, and their spike response function (\cref{fig:1}c) is well described by a logistic function
\begin{equation}
    \label{eq:nuout}
    p(z=1|\el) = \frac{1}{1+\exp(-[\el-u_0]/\alpha)} \,,
\end{equation}
where $u_0$ represents the position of and $\alpha$ the slope at the inflection point.
Note, that changing $\el$ has the same effect as a change of the synaptic input $\isyn$.
In other words, each neuron effectively calculates $p(z=1|\isyn)$, such that the network as a whole can be shown to approximately sample from a Boltzmann distribution $p_\theta\left(\vec{z}\right)=\mathrm{exp}\left[-\epsilon_\theta\left(\vec{z}\right)\right]$ \cite{Petrovici2016} with network energy $\epsilon_\theta\left(\vec{z}\right)=-\sum_{i,j}z_iW_{ij}z_j/2-\sum_{i}z_ib_i$ and  parameters $\theta=\left(\bf{W},\vec{b}\right)$.

One can relate the abstract weights $W_{ij}$ to the physical strength of the synaptic interaction $w_{ij}$ from neuron $i$ to neuron $j$ and the abstract biases $b_i$ to the value of the physical leak potential $\el$ of each neuron.
These two parameter domains are related linearly but have different units.
The mapping between physical neuron and synapse parameters and abstract Boltzmann weights can be gauged by measuring the logistic activation function (\cref{eq:nuout}) with respect to some form of current stimulus.
This relation neglects some dynamic aspects and as such only holds approximately \cite{Petrovici2016}.
This does not restrict the learning scheme applied here.
The probability distribution of physical interest is then the marginal over the hidden space (\cref{fig:1}d,e)
\begin{align}
\label{eq:pv}
  p_\theta\left(\vec{v}\right)&=\frac{1}{Z_\theta}\sum_{\vec{h}}\mathrm{exp}\left[-\epsilon_\theta\left(\vec{z}\right)\right]\;,
\end{align}
which is used to encode the ground state wave function (see \cref{sec:background_C}).
The partition sum $Z_\theta=\sum_{\vec{z}}p_\theta\left(\vec{z}\right)$ ensures proper normalization. 

\subsection{Neuromorphic chip}\label{sec:background_B}

\glsunset{BSS2}
We used the BrainScaleS-2-HICANN-X-v2 physical neuromorphic system \cite{Billaudelle2019} -- in the following abbreviated as \gls{BSS2} -- depicted in \cref{fig:2}a, for all experiments reported in this manuscript. 
\gls{BSS2} is a mixed-signal neuromorphic chip, with 512 \gls{adex} neuron circuits, which we configured to implement current-based \gls{LIF} neurons (see \cref{eq:lif}).
Due to their analog nature, neuron dynamics are 1000 times faster than in their biological counterparts.
Spikes are communicated as digital events which then trigger an analog post-synaptic interaction in downstream neurons.

We employed a routing protocol that forms a freely configurable network of 256 spike sources, combining two neuronal circuits in order to increase their maximum number of presynaptic sources to 256.
64~of these we assigned to the on-chip (noise) spike generators to provide a pool of stochasticity required for sampling \cite{Petrovici2016}.
The full on-chip network structure, including both the sampling network and the noise source allocation, is shown in \Cref{fig:2}b.
The bipartite connection graph is reflected in the block structure of the connection matrix (left part) and the noise sources are randomly assigned from a fixed pool of 32 excitatory and 32 inhibitory sources (right part).
This left us with up to 196 arbitrarily connectable stochastic neurons of which we used a subset to variationally learn the probability distribution representing the ground state wave function of a physical system of interest (see \Cref{fig:1}d,e). 

For each hardware run the \gls{BSS2} chip returns a list of all (output) spike times and associated neuron IDs.
This information combined with the measured $\tauref$ for each neuron is sufficient to reconstruct the network state $\vec{z}(t)$ at any point in time $t$.
We computed the network state at regular intervals, as visualized in \Cref{fig:1}b.
The resulting binary configurations were collected in a histogram as shown in \Cref{fig:1}d and formed an estimate of the steady-state distribution $p(\vec{z})$ of the current network configuration. 
By identifying the neuronal states ($z \in \{0, 1\}$) with the basis states of a qubit system ($\ket{\uparrow}, \ket{\downarrow}$) (see \Cref{fig:1}d),  $p(\vec{z})$ represents the quantum many-body state.
Treating the physical network parameters as variational parameters this representation can be tuned to the ground state on a quantum system, as detailed in the following.

\subsection{Variational algorithm}\label{sec:background_C}

Our goal is to find an approximation of the ground state of a given stoquastic Hamiltonian $H$.
For this we need to determine the parameter set $\theta$ for which our variational anzatz $| \psi_\theta  \rangle$ of the ground state wave function  minimizes the expectation value of the energy:
\begin{equation}
    E_\theta = \langle \psi_\theta|H|\psi_\theta \rangle \,.
    \label{eq:energy}
\end{equation}
The restriction to stoquastic Hamiltonians guarantees that the wave function of the corresponding ground state has non-negative real coefficients in the chosen basis which is the case if all off-diagonal elements of the Hamiltonian are negative \cite{bravyiComplexityStoquasticLocal2007}.
We use this property to directly identify the probability distribution $p_\theta({\vec{v}})$ with the wave function coefficients
\begin{equation}
    |\psi_\theta \rangle = \sum_{\vec{v}} \sqrt{p_\theta({\vec{v}})} | \vec{v} \rangle \,,
\end{equation}
where $p_\theta(\vec{v})$ is estimated by the relative frequency of the occurrence of $\vec{v}$ in the samples generated by the \gls{SNN}.
It resembles the marginal of a Boltzmann distribution (see \cref{eq:pv} and \cref{fig:1}d,e) as discussed above \cite{Petrovici2016}.

We employ a gradient-based minimization of the variational energy $E_\theta$.
Differentiating \cref{eq:energy} with respect to the parameters $\theta = (W_{ij}, b_i)$ results in (see \cref{sec:a_learning_rule} for details)
\begin{align}
    \partial_{W_{ij}} E_\theta 
        = \Big\langle \left(E_{\vec{v}}^\mathrm{loc} - E_\theta\right) z_i z_j \Big\rangle_{p_\theta({\vec{z}})}\label{eq:w_update}
    \\
    \partial_{b_k} E_\theta = \Big \langle \left(E_{\vec{v}}^\mathrm{loc} - E_\theta\right) z_k \Big \rangle_{p_\theta({\vec{z}})} \,, \label{eq:b_update}
\end{align}
where 
\begin{equation}
    E_{\vec{v}}^\mathrm{loc} = \sum_{\vec{v'}} H_{\vec{v}\vec{v'}} \sqrt{p_\theta({\vec{v'}})}/\sqrt{p_\theta({\vec{v}})} \label{eq:Eloc}
\end{equation} 
is the local energy.
Evaluating the local energies requires access to an estimate of the probabilities $p_\theta({\vec{v'}})$ for all states $\vec{v'}$ for which the matrix element $H_{\vec{v}\vec{v'}}=\bra{\vec{v}}H\ket{\vec{v'}}$ of the Hamiltonian is non-vanishing. 
Since no analytical relation between the physical parameters of the spiking network and the abstract parameters $\theta$ of the assumed \gls{RBM} distribution is known, we estimate the probabilities $p_\theta({\vec{v}})$ from samples.
In particular, this means that we need to iterate through the whole collection of generated samples $\{\vec{z}\}\sim p_\theta$ twice.
Once to generate the estimate for $p_\theta({\vec{v'}})$ and once to calculate the averages in \cref{eq:w_update} and \cref{eq:b_update}.

We implement a gradient descent scheme by alternating between the neuromorphic sampling $\{\vec{z}\}\sim p_\theta$ and host-based gradient calculations (see \cref{fig:2}a).
In each iteration the chip is reconfigured according to the gradient given in \cref{eq:w_update} and \cref{eq:b_update} using the ADAM optimizer (\cite{Kingma2014}, see \cref{sec:a_adam} for details).
Each training iteration consists of a single hardware (re)configuration followed by multiple sampling runs (see \cref{tab:phasetrans_params})
of $\SI{0.1}{\second}$ each, which corresponds to $10^5$ independent samples, and subsequent gradient calculation (see \Cref{fig:2}c for relative timings).
We emphasize that only the evaluation part scales with the size of the used network and thereby the represented system, while the sampling time itself is system-size-independent.

In order to track the accuracy of the algorithm, the true ground states $|\psi_0\rangle$ and their exact ground state energy $E_0$ is obtained via numerical diagonalization of the Hamiltonian.
While reaching small energy deviations
\begin{equation}
    \Delta E = \frac{|E - E_0|}{N}
\end{equation}
indicates that the algorithm has converged to the ground state, we also consider the state overlap with the exact ground state, i.e. the quantum infidelity
\begin{equation}
    1-F = 1-|\langle \psi_\theta | \psi_0 \rangle | \,,
\end{equation}
to verify the accuracy of the obtained state representation.
We train for a large number of iterations (typically 1500) keeping track of energy deviations and infidelities.

\subsection{Transverse-field Ising model (TFIM)}\label{sec:background_D}

We test the above algorithm on the 1D \gls{TFIM} whose Hamiltonian consists of nearest-neighbor Ising couplings and a homogeneous transverse field,
\begin{equation}
    H_{\textrm{TFIM}} = -J \sum_{\langle i, j\rangle} \sigma^i_z \sigma^j_z - h \sum_{i=1}^N \sigma^i_x \,,
\end{equation}
where $J$ is the interaction strength, $h$ is the strength of the external field and $\langle i, j \rangle$ signifies nearest neighbor pairs.
Periodic boundary conditions are used such that there is an interaction between spin $1$ and spin $N$.
Furthermore, we consider ferromagnetic interactions where $J>0$ such that alignment of neighboring spins leads to a lower energy.
In this case the Hamiltonian of the \gls{TFIM} in the $z$-basis is stoquastic \cite{bravyiComplexityStoquasticLocal2007}.

In the thermodynamic limit the \gls{TFIM} features a quantum phase transition at the critical point $J = h$ which separates the ordered phase ($h < J$) where the energy is dominated by the spin-spin interactions $\sigma_z^i \sigma_z^{i+1}$ from the disordered phase ($h > J$) where spins increasingly align with the $x$-axis due to the influence of the external field~$\sigma_x^i$.

Thus, the two relevant observables are the magnetization in $x$-direction
\begin{equation}
    \langle \sigma_x \rangle = \frac{\sum_i \langle \sigma_x^i \rangle}{N}
\end{equation}
and the two-point $zz$-correlation function
\begin{equation}
    C_{zz}(d) = \frac{\sum_{i} \langle \sigma_z^i \sigma_z^{i+d} \rangle}{N}\;,
\end{equation}
where $d$ is the distance between spins.

Spin-spin correlations generically fall off exponentially, $C_{zz}(d) \simeq C_0(h) \exp(-d/\xi_{zz}(h))$, while in the vicinity of the critical point this dependence turns into a power law \cite{karlUniversalEquilibriumScaling2017}.
Thus the correlation length $\xi_{zz}$ diverges at the critical point indicating the phase transition point. 
Since we are dealing with finite systems ($N \lesssim 10$) the phase transition point is shifted and the correlation length stays finite, but becomes maximal there.
We also note that in the ferromagnetic phase the ground state is a superposition between two components that are strongly $z$-magnetized in either direction, with an energy gap between symmetric and anti-symmetric superposition that vanishes in the limit of $h/J\to 0$. 
Physically, this leads to spontaneous symmetry breaking in the ferromagnetic phase. 
Interestingly, our \gls{SNN} approximation will show an analogous symmetry breaking effect.

\section{Performance}\label{sec:results}

\begin{figure*}[t]
    \centering
    \begin{tikzpicture}[]
        \begin{scope}
            \node[anchor=north west,yshift=0.cm] (lower) at (0.0, 0.0) {
                \includegraphics[width=0.75\textwidth]{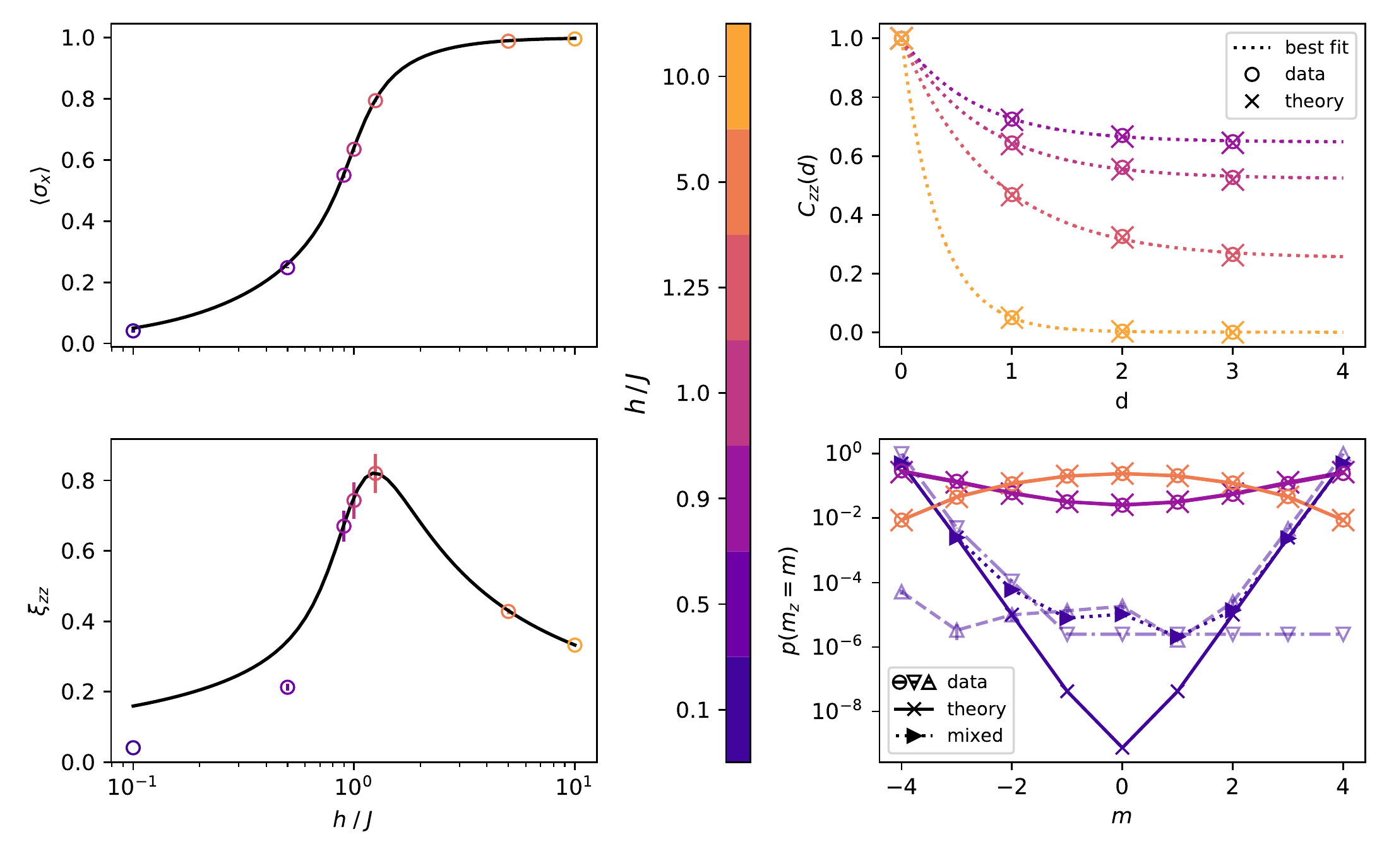}
                };
            \node[font=\bf,xshift=0.5cm,yshift=0.cm] at (lower.north west) {a};
            \node[font=\bf,xshift=7.8cm,yshift=0.cm] at (lower.north west) {b};
            \node[font=\bf,xshift=0.5cm,yshift=-3.9cm] at (lower.north west) {c};
            \node[font=\bf,xshift=7.8cm,yshift=-3.9cm] at (lower.north west) {d};

        \end{scope}
    \end{tikzpicture}
    \caption{
        \textbf{(a)} Average $x$-magnetization $\left<\sigma_x\right>$ of an $N=8$ spin Ising system for different external fields $h / J$.
        Measurement errors are smaller than the marker size and not shown.
        The marker colors identify the field strength $h/J$ in all panels.
        \textbf{(b)} $zz$-correlation for different external fields $h/J \in \{0.1, 0.5, 0.9, 5.0\}$ as function of spin distance $d$.
        An exponential fit (\cref{eq:fit}, dotted line) was applied to the data (circles) which are in agreement with theory (crosses).
        \textbf{(c)} As (a) but for the $zz$-correlation length $\xi_{zz}$.
        Shown error bars are standard deviations over the last 200 training epochs.
        Deviations are observable for small $h$.
        \textbf{(d)} Distribution of observed $z$-magnetization values for different $h$.
        While for $h \in \{1, 5\}$ symmetric distributions are learned, one observes spontaneous symmetry breaking for the lower field value $h = 0.1$.
        In this case, whether the $m>0$ or the $m<0$ component of the ground state is found depends on the choice of initial parameters of the network.
        Averaging over opposite initialisations ($\vartriangle, \triangledown$) results in a good approximation ($\triangleright$, mixed).
        The remaining apparent difference is due to the limited number of samples (where we replaced 0 entries by $10^{-6}$).
        Note also the logarithmic y-axis.
    }
    \label{fig:3}
\end{figure*}
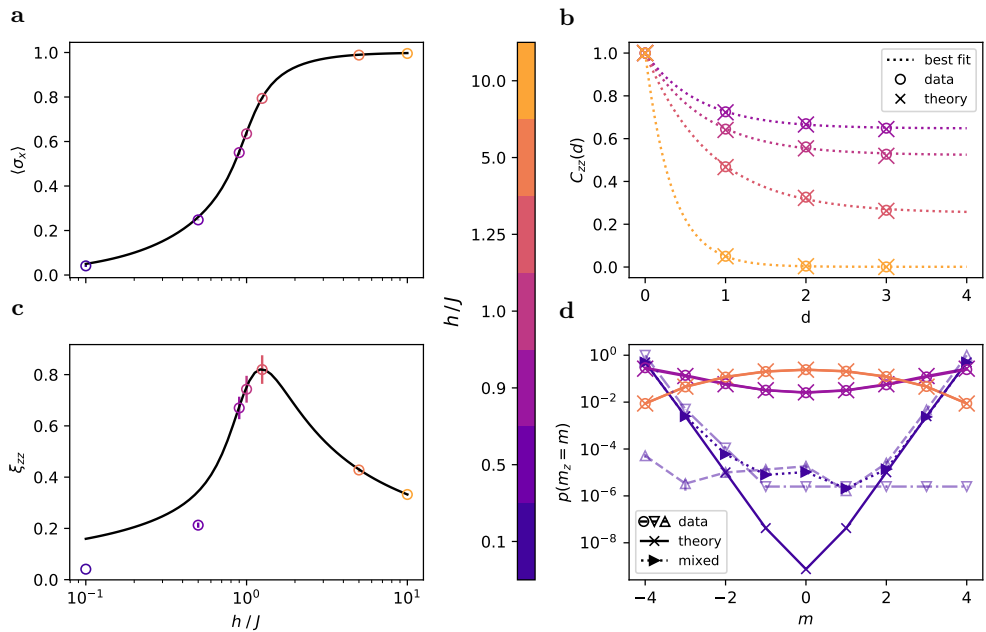

\subsection{Ising phase transition}\label{sec:results_A}

As described above, we trained a generative model using the neuromorphic platform \acrlong{BSS2} to represent ground states of the \gls{TFIM} for a spin chain of size $N=8$ at various transversal field strengths $h \in \{0.1, 0.5, 0.9, 1.0, 1.25, 5, 10 \}$.
The observables shown in \Cref{fig:3} have been obtained through sampling from the learned neuromorphic quantum states.

Overall, we observe very good agreement with the exact solutions for both magnetization (\cref{fig:3}a) and $zz$-correlations (\cref{fig:3}c).
Interestingly, the correlation length systematically deviates for field strengths deeper in the ferromagnetic regime.
As we will show, this happens due to symmetry breaking during the learning process.

In \Cref{fig:3}b the spin-spin correlations in $z$-direction $C_{zz}$ are shown as a function of distance $d$.
The correlation lengths $\xi_{zz}$ are extracted by fitting the data points of each field strength with the following function (shown as dotted lines),
\begin{equation}\label{eq:fit}
    \hat{C}_{zz}(d) = A \exp{(-d/\xi_{zz})} + B
\end{equation}
where the additional parameters $A$ and $B$ account for finite-size effects.
The fit parameters $\xi_{zz}$ and their standard deviations are shown in \Cref{fig:3}c together with the corresponding theoretical values (solid line).
We observe that the correlation length has a maximum at $h/J \approx 1.25$ marking the phase transition point and closely matching the theoretical prediction.
While for $h/J \geq 0.9$ the results agree well with the exact values for both observables, $\langle \sigma_x\rangle$ and $\xi_{zz}$, at $h/J \in \{0.1,0.5\}$ the correlation length is significantly underestimated.

To illustrate the origin of this deviation, we show the probabilities for finding the system in a state with $z$-magnetization $m$ (half of the difference between the number of up- and down-spins in $\vec{v}$) in \Cref{fig:3}d. 
This reveals that instead of the symmetrical ground state distribution which is learned correctly for $h \geq 0.9$, the symmetry is broken for low field values.
For $h=0.1$ it is shown that two different ground states with all spins up or down can be reached.
The average of these two distributions (dotted line) is a good approximation to the symmetric distribution.

Such spontaneous symmetry breaking happens physically whenever the ground state of the system is (near-) degenerate, since any small perturbation of the system will break the symmetry and collapse the macroscopic superposition into one of its components.
That is the exact ground state becomes harder to prepare for $h \rightarrow 0$ as it then increasingly approaches a superposition of the two extreme configurations $\ket{{\downarrow}}^{\otimes N}$ and $\ket{{\uparrow}}^{\otimes N}$.
In a way, we see the same behavior reproduced by the neuromorphic device.
This is because in terms of \gls{SNN} dynamics, such a distribution requires both highly synchronous activity and synchronous inactivity.
Achieving such a behavior requires distributions with strong local minima, making it hard for any \gls{MCMC} method to escape.
This so-called mixing problem already manifested itself in the increased need for samples at $h/J = 0.9$ in order to well represent the symmetric ground state.
The points $h/J \in \{0.1,0.5\}$, are even deeper in the ferromagnetic phase which made learning these highly entangled states prohibitively hard with our static stochasticity system.

\subsection{Dependence on system size}\label{sec:results_B}

\begin{figure*}[t]
    \centering
    \begin{tikzpicture}[]
        \begin{scope}
            \node[anchor=north west] (figure) at (0.0, 0.0) {
                \includegraphics[width=0.75\textwidth]{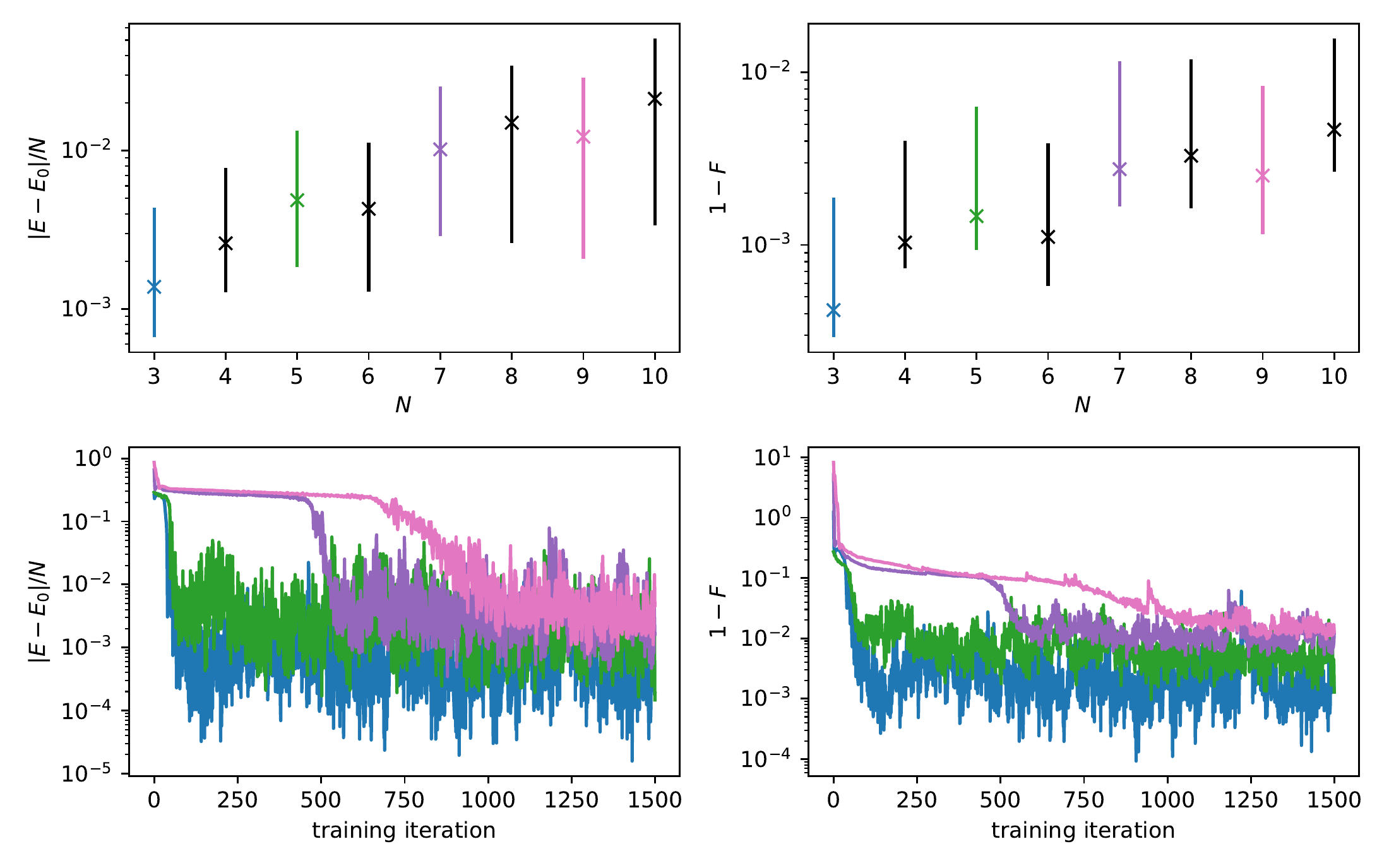}
                };

            \node[font=\bf,xshift=0.5cm,yshift=-0.2cm] at (figure.north west) {a};
            \node[font=\bf,xshift=7cm,yshift=-0.2cm] at (figure.north west) {b};
            \node[font=\bf,xshift=0.5cm,yshift=-4.4cm] at (figure.north west) {c};
            \node[font=\bf,xshift=7cm,yshift=-4.4cm] at (figure.north west) {d};

        \end{scope}
    \end{tikzpicture}
    \caption{
        Performance at $h/J=1$ as a function of system size.
        \textbf{(a)},\textbf{(b)} Relative energy mismatch and infidelity between the learned and exact ground state increases at fixed number of hidden units.
        We report median values and the 15- and 85-percentiles over the last 200 iterations as error bars.
        \textbf{(c)},\textbf{(d)} Evolution of the approximation quality during learning.
    }
    \label{fig:4}
\end{figure*}
In order to assess the scalability of our approach we studied its performance for different sizes of the quantum system.
In the experiment shown in \Cref{fig:4} the number of spins $N$ is increased from $N=3$ to $N=10$ for the critical point $h/J=1$.
For details of the used network parameters and sample sizes, see \cref{sec:a_figsi}. 
Note that the spiking neural network has less parameters than the number of wave function coefficients for $N=9$ and $N=10$.

Overall, a quantum fidelity greater than $99\%$ can be achieved and even up to $99.9\%$ for systems $N\leq6$ (\Cref{fig:4}b).
Since the fidelity imposes an upper bound on the errors of any possible expectation values, good agreement of the learned observables is guaranteed.

\Cref{fig:4}c,d show exemplary learning curves of energy error and fidelity as a function of the training iteration.
While the learning curves converge quickly for small system sizes, it takes progressively longer to reach good metrics for larger systems, with intermediate regimes of very slow improvement.
This behavior is well-studied in the machine learning literature and related to the high number of saddle-points in the parameter space \cite{dauphin2014identifying}.
It should be noted that these plateaus are not observed for $h/J=10$, where the ground state distribution becomes more uniform, which is easy to reach by gradient descent independent of the initial conditions.

\section{Limitations} \label{sec:limitations}

For system sizes above $N=10$, we observed a significant drop in the representational power of our neuromorphic implementation.
In the following, we investigate possible causes in order to ascertain the scalability of our approach.
In particular, we consider (i) limited hidden layer size, (ii) limited network parameter range, (iii) finite network parameter resolution, (iv) non-optimal choice of the learning rate and (v) deviations of the substrate from the theoretically assumed dynamics.
In the following subsections we thoroughly study the impact of substrate induced limitations on the performance of our method.
As we show below, we found that our approach is not limited in principle, but rather in practice by the parameter stability of the used substrate.
This also suggests a clear strategy for scaling to larger system sizes.

\subsection{Hidden layer size}\label{sec:results_C}

In order to assess the required number of hidden units for a good variational representation depending on the system size, we have performed a grid search over $(N, N_h) = (\{3,\ldots, 8\}, \{5,10,20\} )$ drawing \mbox{$N_\mathrm{sample} = 2\cdot10^5$} samples for $1500$ training iterations each.

\begin{figure*}[t]
    \centering
    \begin{tikzpicture}[]
        \begin{scope}
            \node[anchor=north west] (figure) at (0.0, 0.0) {
                \includegraphics[width=0.75\textwidth]{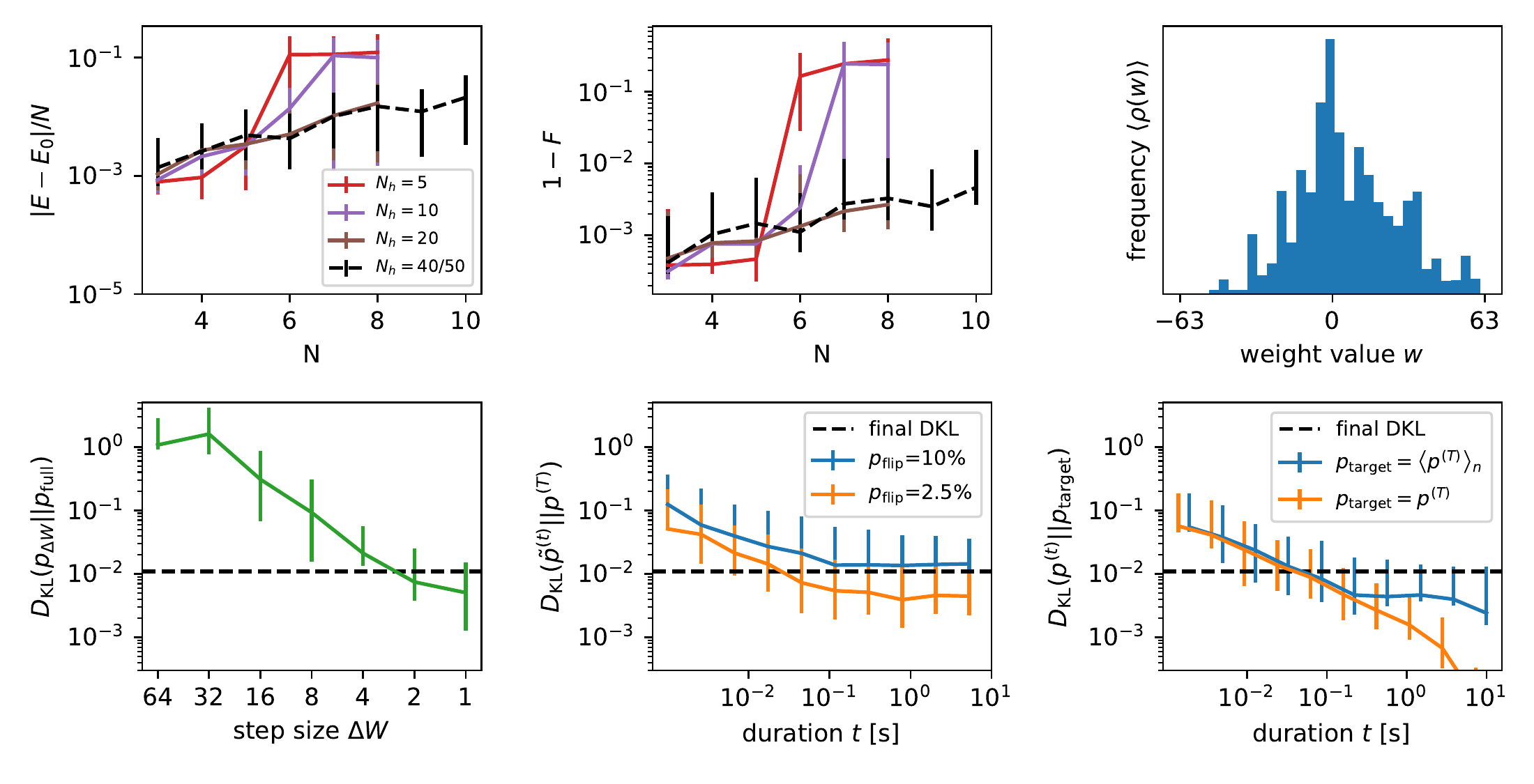}
                };

            \node[font=\bf,xshift=0.5cm,yshift=-0.2cm] at (figure.north west) {a};
            \node[font=\bf,xshift=5cm,yshift=-0.2cm] at (figure.north west) {b};
            \node[font=\bf,xshift=9.5cm,yshift=-0.2cm] at (figure.north west) {c};
            \node[font=\bf,xshift=0.5cm,yshift=-3.55cm] at (figure.north west) {d};
            \node[font=\bf,xshift=5cm,yshift=-3.55cm] at (figure.north west) {e};
            \node[font=\bf,xshift=9.5cm,yshift=-3.55cm] at (figure.north west) {f};
        \end{scope}
    \end{tikzpicture}
    \caption{
        \textbf{(a)},\textbf{(b)}~Approximation quality as a function of the hidden layer size, analogous to \Cref{fig:4}a,b.
        Small hidden layers can limit the fidelity of the learned state, especially for larger systems.
        For the system sizes used here ($N<10$) hidden layer sizes of $N_h=20$ have proven to be sufficient.
        \textbf{(c)}~Weight distribution accumulated over the final $200$ epochs for $N=8$, $N_h=20$.
        The weights are not clipped significantly by the limited range $-63 \leq w \leq +63$.
        \textbf{(d)}~Effect of the weight resolution:
        \gls{DKL} between the full 7-bit distribution and a coarse grained one as a function of the smallest possible weight step $\Delta w$.
        For comparison: A successfully trained system with $N=8$ and $N_h=20$ reached a final \gls{DKL}$\approx10^{-2}$ (dashed horizontal line, also in (e) and (f)).
        \textbf{(e)}~Comparison between a reference distribution $p^{(T)}$ and a distribution perturbed by a "pseudo weight update".
        We show the \gls{DKL} between these distributions as a function of hardware execution time $t<T$ used for sampling the perturbed distributions $\tilde{p}^{(t)}$.
        \textbf{(f)}~Convergence behavior for a static configuration:
        Comparing to the final distribution of a single run (orange) we observe the ideal $1/N_\mathrm{sample}$ behavior.
        Convergence towards an average distribution $\langle p^{(T)} \rangle_n$ over multiple runs stops at a sampling time of about $\SI{0.2}{\second}$ (blue).
        Note, for visibility reasons we plot alternative times for the different experiments.
    }
    \label{fig:5}
\end{figure*}
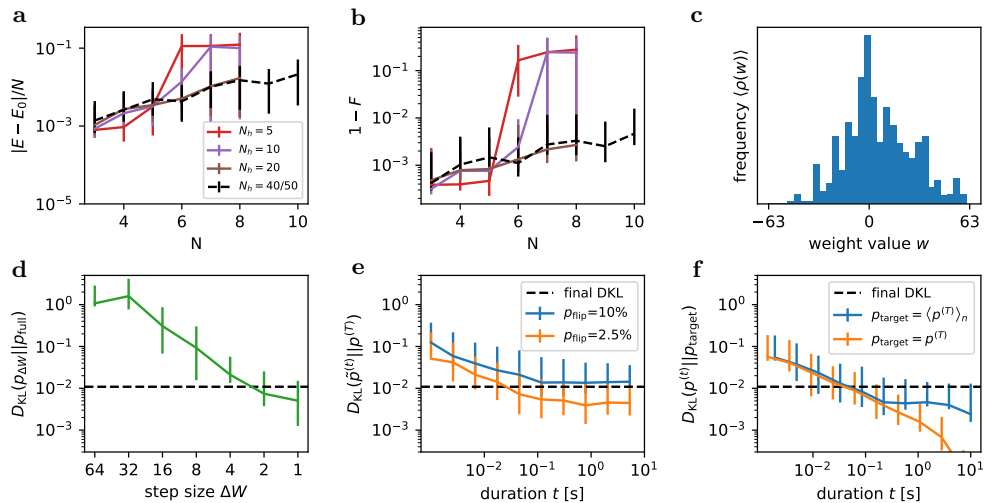

The results are shown in \Cref{fig:5}a,b in terms of median energy error per spin and infidelity of the state representation averaged over the last $200$ training iterations.
While $N_h=5$ (red line) is sufficient to accurately describe the systems for $N < 6$, both energy error and infidelity increase sharply for larger $N$.
Increasing the number of hidden neurons to $N_h=10$ (purple line) allows us to obtain accurate ground state representations up to $N=6$ with $F> 99\%$, while $N_h\geq 20$ (brown line) is required for $N \in \{7,8\}$.

One might expect that using more hidden units could decrease the slope of the curve further, also bringing the large systems above $99.9\%$ fidelity.
However, comparing with $N_h=40$ (black dashed line, from \Cref{fig:4}a,b), there is no significant difference in either energy error or fidelity, suggesting that model capacity is not the dominating limitation.
While the system is indeed overparameterized in the sense that there are more variational parameters than wave function coefficients, the physical network parameters can only be controlled with finite precision, and within a finite range.
These bounds may limit the representational power of the ansatz, as we discuss in the next two sections.

\subsection{Weight range}\label{sec:weight_distribution}

The strongest realizable weights on \gls{BSS2} are represented by the digital values $w_\mathrm{max} = \pm 63$.
\Cref{fig:5}c shows a typical weight distribution accumulated over the last 200 training iterations (for $N=8$ and $N_h=20$).
The distribution is peaked around zero with roughly symmetrical tails.
However, there is no significant occupancy of the outermost weight values, hence clipping beyond the edges of $\pm63$ should have no effect.
We conclude that the chosen
weight range is sufficient and does not restrict the achievable representation accuracy. 

\subsection{Weight resolution}\label{sec:results_D}

On the \gls{BSS2} system, the synaptic connections 
are implemented by two 6-bit configurable circuits, one for the excitatory ($w_{ij}>0$) and one for the inhibitory ($w_{ij}<0$) part of the synaptic connectome.
We therefore used two physical synapses to form a logical synapse which gives an additional bit for the sign (see \cref{sec:a_bssparams} for details).

A lower parameter resolution leads to a more coarse-grained space of representable distributions.
To assess whether this has a detrimental impact, and hence is a limiting factor for learning performance, we conducted an experiment where we randomly initialized the network weights ($N_v=8, N_h=20$) drawing from a uniform distribution $w_{ij} \sim \mathcal{U}(-63, 63)$.
The neuron biases were set to the center (i.e., half the maximum spike rate) of their respective activation functions (cf. \cref{fig:1}c). 
We then artificially reduced the resolution of the weights, and thus of the distribution, by defining a grid centered at zero and with a minimum step size $\Delta w > 1$ between two allowed weight values.
We compare distributions sampled using the full parameter resolution  $p_\text{full} = p_{\Delta w=1}$
to the distributions $p_{\Delta w}$ obtained by rounding the weights to the low resolution grids with step sizes $\Delta w$.
We quantify the distance between these distributions by the \acrlong{DKL}
\begin{equation}
    D_\mathrm{KL}(p\Vert q) = -\sum_v p(v) \log(p(v)/q(v))
\end{equation}
such that $p = p_\text{full}$ and $q = p_{\Delta w}$.
For every $\Delta w$ we repeat 10 sampling experiments each of duration $T=\SI{0.1}{\second}$.

As \Cref{fig:5}d shows, we find a quick decrease in \gls{DKL} as the step size shrinks, which, however, plateaus for \mbox{$\Delta W \leq 2$}.
The achieved \gls{DKL} for $\Delta w \in \{1, 2\}$ is consistent with the typical final $D_\text{KL} \simeq 10^{-2}$ (dashed line) for trained networks of the same size. 
Therefore, we conclude that the limited parameter precision also does not explain the saturation in the observed learning performance.

Note that the network observed here had ample representational power for the system size (cf. \cref{fig:4}).
It may be that a more significant effect could be observed for smaller hidden layers.
Furthermore, no training was performed in order to isolate the effect of finite weight resolution on the accuracy of the sampled distribution.

\subsection{Learning rate}\label{sec:learning_rate}

In all our ground state learning experiments we used a learning rate decay (see \cref{sec:a_adam}) to facilitate the descent into minima of the energy landscape.
Additionally, towards the end of the training the gradients become small thereby also shrinking the weight updates.
At late stages of the training we typically observe changes in $2-3\%$ of the individual discrete weights.
A potential limitation to the achievable convergence is a still too high learning rate at the end of the training which prevents precise descent into local minima.
To test whether this is the case, we perturbed a reference distribution with a "pseudo update" and observed the size of the resulting deviation measured by the \gls{DKL}.

In particular, we again initialized our system with a uniformly random distributed weight matrix \mbox{$w_{ij}\sim \mathcal{U}(-62,62)$} and collected samples from it over a period of $T=\SI{10}{\second}$ -- significantly longer than needed for convergence $\approx\SI{0.1}{\second}$.
This defined our reference distribution $p^{(T)}$.
We then simulated a weight update by changing a fraction $p_\mathrm{flip}$ of the weight parameters $w_{ij}$ by $\pm1$ and again sampled from the modified distribution.
This defined a perturbed distribution $\tilde{p}^{(T)}$.
Using only samples up to some time $t<T$ defined a series of perturbed distributions $\tilde{p}^{(t)}$.
In \cref{fig:5}e we show the evolution of the resulting \gls{DKL} between these perturbed distributions and the reference distribution.

We observe that the \gls{DKL} (green curve) decreases quickly until around $t = \SI{0.1}{\second}$ after which it saturates due to the distortion induced by the random weight changes.
$p_\mathrm{flip}=10\%$ was chosen such that the final \gls{DKL} corresponds to the observed final \gls{DKL}s during training (dashed horizontal line).
On the other hand, for a value of $p_\mathrm{flip}=2.5\%$ we observe that a better approximation is reached.
Since we observed 2-3\% weight flips per learning update at the late stages of the actual training, this result indicates that the ground state search is not limited by a too large learning rate.

\subsection{Temporal stability of the substrate}\label{sec:results_E}

The key feature of the \gls{BSS2} system -- and the main catalyst of its speed and efficiency -- is the analog nature of its neuro-synaptic dynamics.
However, its direct benefits for our approach come with a number of specific challenges that do not appear in digital devices or simulations, such as a certain amount of component diversity, as shown in \cref{fig:1}c.
While this particular effect is automatically corrected for during learning, other phenomena are more subtle and difficult to compensate.
An immanent property of analog components is the presence of small instabilities and drifts in their parameters \cite{Schemmel2010,pfeil2013six,Schmitt2017}.
In this section we study the impact of such effects on the sampling and learning performance by conducting long-duration sampling experiments of $T=\SI{10}{\second}$ (initialization as in \cref{sec:results_D}).

First, we compare the convergence during a single run by measuring $\DKL(\tilde{p}^{(t)} \Vert p^{(T)})$ (orange line in \cref{fig:5}f).
Here, by construction, convergence to zero is assured and we see the expected $1/N_\text{sample}$ behavior of Monte Carlo sampling.
Our aim was to test the reproducibility and stability of the sampling procedure over multiple iterations and reconfigurations for a fixed parameter set $\theta$.
In a second experiment, we therefore repeated the sampling procedure for $T=\SI{10}{\second}$ for $n=30$ times and averaged the resulting distributions.
We then compared the observed distributions $p^{(t)}$ to the average target distribution $p_\mathrm{target}=\left<p^{(T)}\right>_n$.
Initially, the \gls{DKL} gradually decreased as more samples were gathered (\cref{fig:5}f).
However, beyond $\SI{200}{\milli\second}$ the \gls{DKL} saturated.
This shows that even a repetition with the exact same configuration of network parameters $\theta$ samples from a slightly different distribution than the original one $p_\theta$.
This, in turn, indicates that the parameters of the physical system do not stay constant over the duration of an entire experiment.

The time scale of variability observed above is significantly shorter than the total duration of both training and evaluation, each of which covered at least 200 epochs of $\SI{100}{\milli\second}$.
We thus conclude that the temporal variability of the analog parameters represents the main limiting factor for the fidelity of our approach on \gls{BSS2}.
For larger system sizes, where more samples are required to obtain precise gradient estimates, this effect becomes increasingly severe and thus causes the observed drop in the representational power of our neuromorphic implementation.
Understanding this limitation points directly to possible mitigation strategies, which we address in the discusion below.

\section{Discussion}
\label{sec:conclusion}

In summary, we have presented a demonstration of neuromorphic ground state search for quantum spin systems.
We have designed a variational algorithm suitable for implementation in the mixed-signal \gls{BSS2} system which enables fast spike-based sampling in an inherently parallel fashion and independent of the network size.
These advantages could provide significant speedups for the emulation of large networks or quantum spin systems.
For this reason we have tested the scalability of our approach, thereby expanding previous work by Czischek et al. \cite{czischekSpikingNeuromorphicChip2021} from representing small entangled states to larger quantum spin systems of up to $N=10$ spins.
Furthermore, we have analyzed the \gls{TFIM} phase transition and found excellent agreement with exact solutions.
In the ferromagnetic regime we observed symmetry breaking in the \gls{SNN} activity reflecting the tendency of the quantum spin system towards spontaneous order.

For systems with $N>10$, the reachable approximation quality decreased sharply.
By systematically studying potential limiting factors, we were able to exclude several possible causes of this degradation, namely the limited number of hidden neurons, finite weight range and resolution, as well as non-optimal learning rate.
Moreover, we found that the currently available parameter stability on \gls{BSS2} leads to a limited accuracy of gradients and thus represents the main technical obstacle to be overcome for further improving the approximation quality at large system sizes.
A second, algorithmic limitation of our learning scheme is the requirement of the precise knowledge of $p(\vec{v}')$ for all non-zero observed $p(\vec{v})$ that are connected by a non-zero $H_{\vec{v}\vec{v'}}$.

Neither the technical nor the algorithmic challenges are fundamental roadblocks for using neurmorphic hardware for variational learning of quantum states and will be addressed in future research.
Since \gls{BSS2} was developed as a multi-purpose research system, its capabilities were not optimized for spike-based sampling.
Advancements in the development of \gls{BSS2} and other neuromorphic hardware platforms \cite{Roy2019} will alleviate technical issues and introduce new capabilities and tools.
For analog, and in particular accelerated platforms, parameter variability over typical experiment durations of tens to thousands of seconds can be greatly reduced.
On the other hand, using purely digital neuromorphic chips such as ODIN \cite{frenkel20180} or Loihi \cite{Davies2018} would circumvent the instabilities of an analog system and thus permit scaling to larger quantum system sizes.
While this might come at the cost of losing some of the analog advantages of \gls{BSS2}, mainly with respect to speed and energy efficiency, it will likely still outperform more conventional, CPU/GPU-based solutions \cite{goltz2021fast}.
In either scenario, improved control and readout of the neuromorphic substrate could also allow the direct calculation of Boltzmann factors from the weight and bias parameters.
This would enable the efficient computation of local energies [\cref{eq:Eloc}] and thus solve the problem of having to densely sample the visible distribution \cite{Carleo2017}.

For small transverse fields we observed a symmetry breaking during the training.
The parity symmetry corresponding to a global spin flip required two differently initialized training runs to be reproduced deep in the ferromagnetic phase.
The root cause is the near-degeneracy of the two highly synchronous states (all active, all inactive).
This corresponds to the well-known mixing problem for which spike-based solutions have been proposed \cite{korcsak2020cortical,leng2018spiking} which would be amenable to a neuromorphic implementation and should allow a faithful representation of such distributions without the need for re-initialization.
From an algorithmic perspective one could also enforce this symmetry by supplementing the generated sample sets with the corresponding spin-flipped configuration for each sample generated by the network.
This technique can be employed to enforce any given symmetry of the physical model \cite{Choo2018,Bukov2021,Nomura2021}.

Another promising idea for scalable algorithms is the use of local learning rules that only involve connected neuron pairs since most modern neuromorphic platforms support local on-chip learning.
An example for training \gls{RBM}s with a local learning rule is contrastive divergence \cite{hintonPracticalGuideTraining2012a}, for which an event-driven \gls{SNN} version has been proposed \cite{neftciEventdrivenContrastiveDivergence2014a}.
Additionally, such generative networks can be further fine-tuned using error backpropagation, which, in turn, can be approximated by local learning rules \cite{whittingtonApproximationErrorBackpropagation2017,sacramento2018dendritic,craftonDirectFeedbackAlignment2019a, leeSpikeTrainLevelDirect2020,haider2021latent}, including spike-based variants already demonstrated on \gls{BSS2} \cite{Billaudelle2019,goltz2021fast}.
The question of how to translate these local update schemes to variational ground state learning is left as an important direction for future research.

Finally, algorithmic improvements could be enabled by novel encodings of NQS with \gls{SNN}s.
A straightforward idea for encoding not only the amplitudes, but also phases of the wavefunction would be to use additional output units or even a second network like in \cite{Torlai2018}.
Phasor networks represent another possible avenue for encoding complex numbers with \gls{SNN}s.
It was shown that these networks, which consist of resonate-and-fire neurons with complex dynamical variables, can be implemented by integrate-and-fire \gls{SNN}s and can robustly leverage spike-timing codes  \cite{fradyRobustComputationRhythmic2019}.
If successful, these approaches to representing complex values in \gls{SNN}s could enable the extension of the presented variational method to non-stoquastic systems.

\begin{acknowledgments}
We thank G.\ Carleo, S.\ Czischek and T.\ Gasenzer for helpful
discussions.
We thank the Electronic Vision(s) group, in particular S. Billaudelle, B. Cramer, J. Schemmel, E. C. M{\"u}ller, C. Mauch, Y. Stradmann, P. Spilger, J. Weis, and A. Emmel, for maintaining and providing access to the BrainScaleS-2 system and for technical support.
This work is supported by the Deutsche Forschungsgemeinschaft (DFG, German Research Foundation) under Germany's Excellence Strategy EXC 2181/1 -- 390900948 (the Heidelberg STRUCTURES Excellence Cluster), by the DFG -- project-ID 273811115 -- SFB 1225 (ISOQUANT), by the European Union 7th and Horizon-2020 Framework Programmes, through the ERC Advanced Grant EntangleGen (Project-ID 694561) and under grant agreements 785907, 945539 (HBP), and by the Manfred St{\"a}rk Foundation.
\end{acknowledgments}

\section*{Methods}
\renewcommand{\appendixname}{Methods}
\appendix

\section{Derivation of the learning rule}\label[app]{sec:a_learning_rule}

We calculate the derivative of the variational energy with respect to a weight $W_{ij}$ of the network assuming a stoquastic Hamiltonian $H$ and the normalized state representation $|\psi\rangle = \sum_{\vec{v}} \sqrt{p(\vec{v})} | \vec{v} \rangle $:
\begin{align}
\partial_{W_{ij}} E_\theta &= \partial_{W_{ij}} \sum_{\vec{v}\vec{v'}} \sqrt{p({\vec{v}}) p({\vec{v'}})} H_{\vec{v} \vec{v'}} \\\label{eq:der_evar1}
&= \sum_{\vec{v}\vec{v'}} \sqrt{\frac{p({\vec{v'}})}{p({\vec{v}})}} H_{\vec{v} \vec{v'}} \partial_{W_{ij}} p({\vec{v}}) \\\label{eq:der_evar2}
&= \sum_{\vec{v}\vec{v'}} \sqrt{\frac{p({\vec{v}})}{p({\vec{v'}})}} H_{\vec{v} \vec{v'}} \nonumber \\ &\cdot \left( \sum_{\vec{h}} z_i z_j p({\vec{v'}}) - p({\vec{v}}) \sum_{\vec{z}'} z'_i z'_j p({\vec{z}'}) \right) \\
&= \sum_{\vec{v}\vec{h}} E_{\vec{i}}^\mathrm{loc} z_i z_j p(\vec{z}) - \sum_{\vec{i}} E_{\vec{v}}^\mathrm{loc} p({\vec{v}}) \sum_{\vec{z}'} z'_i z'_j p({\vec{z}'}) \label{eq:qintro} \\ 
&= \sum_{\vec{z}} \left(E_{\vec{v}}^\mathrm{loc} - E_\theta\right) z_i z_j p({\vec{z})} \label{eq:eintro} \\
&= \Big\langle \left(E_{\vec{v}}^\mathrm{loc} - E_\theta\right) z_i z_j \Big\rangle_{p({\vec{z}})} \,.
\end{align}
From equation~(\ref{eq:der_evar1}) to (\ref{eq:der_evar2}) we have used the symmetry of the Hamiltonian.
In (\ref{eq:qintro}) the local energy $E_{\vec{v}}^\mathrm{loc} = \sum_{\vec{v'}} H_{\vec{v}\vec{v'}} \sqrt{p({\vec{v'}})}/\sqrt{p({\vec{v}})}$ is introduced and the variational energy appears in (\ref{eq:eintro}) due to the relation $E_\theta = \sum_{\vec{v}} E_{\vec{v}}^\mathrm{loc} p(\vec{v})$.

To deal with the numerical problem of vanishing entries in $p_\theta(\vec{v})$ a small parameter $\epsilon$ is added to it, essentially introducing a bias towards a uniform distribution.
The local energy thus reads $E_{\vec{v}}^\mathrm{loc} = \sum_{\vec{v}} H_{\vec{v}\vec{v'}} \sqrt{p_{\vec{v'}} + \epsilon}/ \sqrt{p_{\vec{v}} + \epsilon} $ where $\epsilon =10^{-12}$ was used throughout.

With the above derivation the gradient of the BM, $\nabla_\theta E_\theta \equiv \Delta \theta_\text{BM}$, can be estimated as sample average.
For the learning scheme we assume that the variational energy gradient with respect to the \gls{BSS2} hardware parameters, $\Delta \theta_\text{BSS}$, is well approximated by the analogous computation over hardware samples.

\section{Adaptive momentum optimization}\label[app]{sec:a_adam}

Since the gradient only provides local guidance, it is advisable to scale its components according to the roughness of the cost landscape.
An adaptive step size decay probes the cost surface at increasing resolution as the training progresses and bounds the number of steps that need to be computed to reach convergence.
We typically employed an exponentially decaying step size $\alpha(t+1) = \alpha(t) \cdot \gamma_\mathrm{lr}$ such that $1/(1 -\gamma_\mathrm{lr})$ sets a time scale of required optimization steps.
We found the values $\alpha(1) = 1$, $\gamma_\mathrm{lr} = 0.999$ to work well in practice.

In addition to the fixed step size decay, we employed the Adam scheme \cite{Kingma2014} which combines momentum with an adaptive learning rate which is chosen for each network parameter individually.
It is a first-order method that estimates mean, $m(t)$, and variance, $v(t)$, of the gradient by exponential running averages with respective decay rates $\beta_1$ and $\beta_2$:
\begin{align}
m(t+1) &\leftarrow \frac{\beta_1}{1 - \beta_1^t} m(t) + \frac{1 - \beta_1}{1 - \beta_1^t} \Delta \theta_\text{BSS}(t) \\
v(t+1) &\leftarrow  \frac{\beta_2}{1 - \beta_2^t} v(t) +  \frac{1 - \beta_2}{1 - \beta_2^t}  \Delta \theta^2_\text{BSS}(t)
\end{align}
where $\Delta \theta^2_\text{BSS}(t)$ is the component-wise square of the gradient.

The parameters are updated according to the inverted relative error of the gradient where $m(t)$ acts as a momentum and $v(t)$ modifies the learning rate
\begin{equation}
\theta_\text{BSS}(t+1) \leftarrow \theta_\text{BSS}(t) - \eta(t) \frac{m(t)}{\sqrt{v(t)} + \epsilon_\mathrm{Adam}} \,.
\end{equation}
The small parameter $\epsilon_\mathrm{Adam}$ is required for regularization purposes.
Since $|\Delta \theta(t)/\sqrt{v(t)}| \leq 1$ the update implicitly adapts the step sizes based on the signal-to-noise ratio of the derivatives.
The canonical hyperparameters for Adam are used: $\beta_1 = 0.9$, $\beta_2 = 0.999$, $\epsilon_\mathrm{Adam} = 10^{-8}$.

\section{Leaky integrate-and-fire neurons}\label[app]{sec:a_lif}

The \gls{LIF} neuron model belongs to the family of continuous spiking neuron models \cite{gerstnerSpikingNeuronModels2002}.
The neuron's membrane is modeled as a capacitor with capacitance $C_m$.
It can be charged by the synaptic current stimulus $I^\mathrm{syn}(t)$ while it is constantly discharged across a leak conductance $g_l$.

According to Kirchhoff's laws the voltage $u$ across the capacitance is described by
\begin{equation}
C_m \frac{\text{d} u(t)}{\text{d} t} = g_l(V_l - u(t)) + I^\mathrm{syn}(t) \,.
\end{equation}
The potential $V_l$ plays the role of the resting state which is, in the absence of external input,  approached on the time scale of the circuit $\tau_m = C_m/g_l$.

The spike mechanism is triggered when the membrane potential crosses a threshold $V_\text{thresh}$ from below:
\begin{equation}
u(t_{\mathrm{spike}}) = V_\text{thresh} \land u'(t_{\mathrm{spike}}) > 0 \,.
\end{equation}
After the spike has been fired, the membrane potential is clamped to a reset value during the absolute refractory period $\tau_{\mathrm{ref}}$:
\begin{equation}
u(t_{\mathrm{spike}} \leq t \leq t_{\mathrm{spike}} + \tau_{\mathrm{ref}}) = V_\text{reset} \,.
\end{equation}

\gls{BSS2} implements current-based synapses in which case synaptic weights carry the unit of current.
The spike input of neuron $j$ is determined by the exponential synaptic kernel $\kappa(t) = \Theta(t) \exp\left(-t/\tau_\mathrm{syn}\right)$ convolved with spike trains of presynaptic neurons $S^i(t) = \sum_{t_s} \delta(t - t_s^i)$:
\begin{equation}
I^\mathrm{syn}_j(t) = \sum_i w_{ij} (S^i \star \kappa)(t) = \sum_{i} w_{ij} \sum_{t_s^i} \kappa(t - t_s) \,.
\end{equation}
The influence of spikes thus decays with the time scale  $\tau_\mathrm{syn}$.

\begin{figure}
    \centering
    \includegraphics[width=0.5\columnwidth]{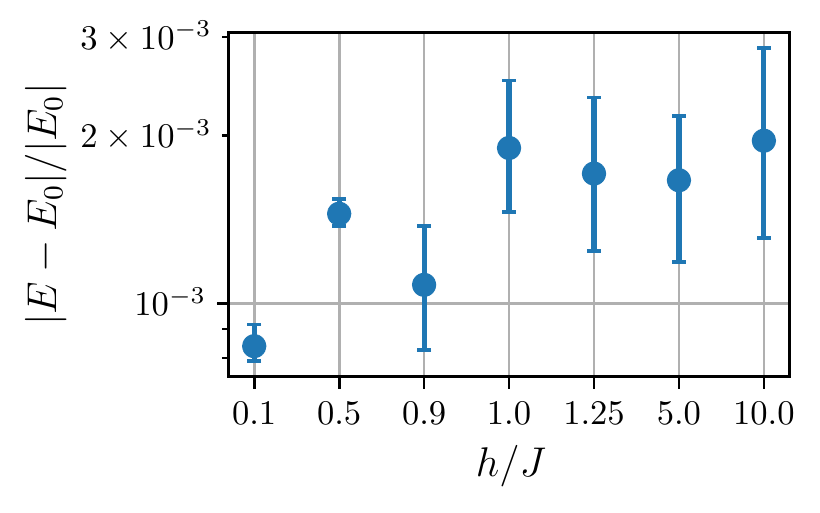}
    \includegraphics[width=0.46\columnwidth]{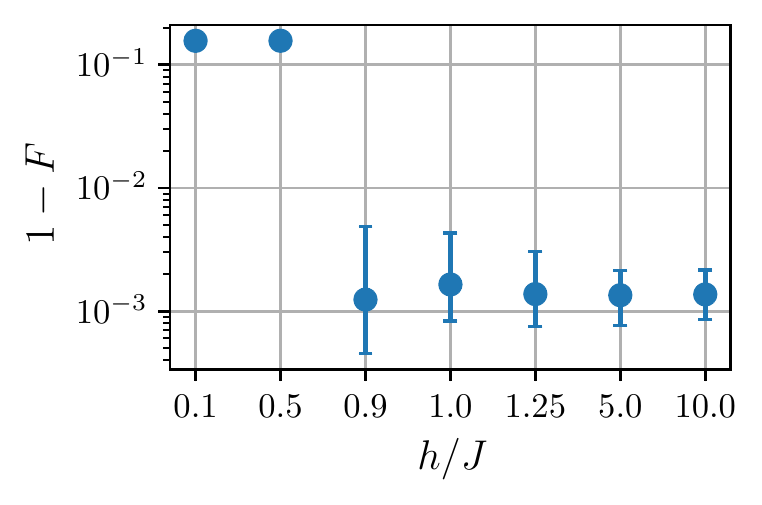}
    \caption{Relative energy error (left) and infidelity (right) as function of $h/J$. 
    Note that the field values are not spaced equidistantly. The network parameters used in Figs.~\ref{fig:si_fig1_energy_infidelity} and \ref{fig:si_symbreak} are as in Fig.~\ref{fig:1}.
    }
    \label{fig:si_fig1_energy_infidelity}
\end{figure}

\begin{figure}
    \centering
    \includegraphics[width=0.7\columnwidth]{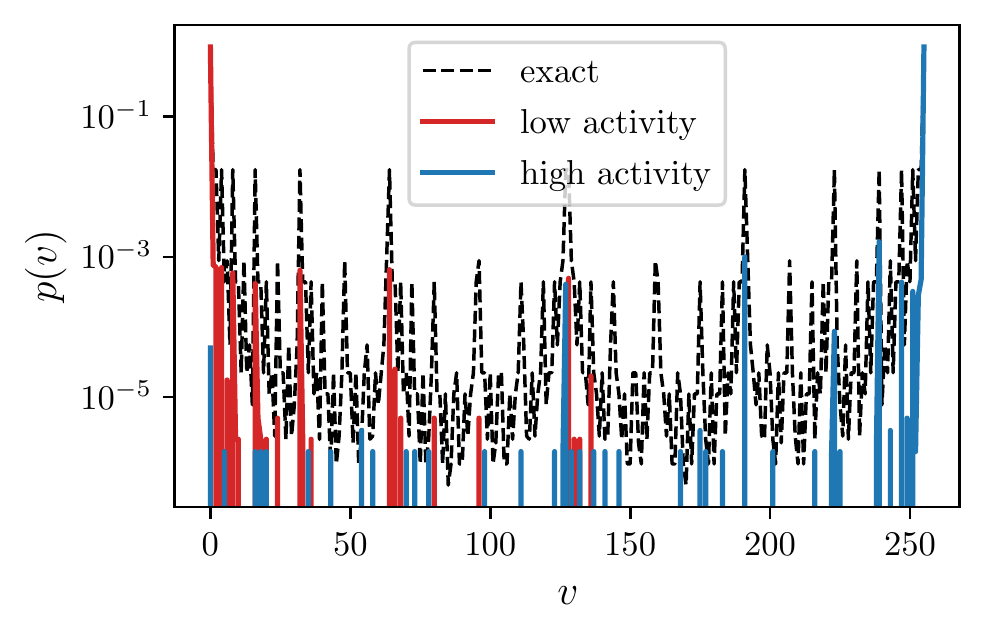}
    \caption{Symmetry breaking at $h/J = 0.1$.
    Learned probability distribution over visible neuron configurations $v$ corresponding to basis states of the spin system (solid) compared to the exact ground state distribution (dashed).
    Standard initialization (blue) favours a high activity state, while an initial negative bias offset on all neurons (red) results in final state with low network activity.
    }
    \label{fig:si_symbreak}
\end{figure}

\section{Supplementary analysis of symmetry breaking}\label[app]{sec:a_symmetry_breaking}

In the experiments shown in \Cref{fig:3} we saw that the symmetry of the ground state was broken for $h \in \{0.1, 0.5\}$ in favor of "spin up" or simultaneous firing of all visible neurons.
This bias for the high activity state might be due to the exponential synaptic kernel's influence extending beyond the refractory period. 

Figures \ref{fig:si_fig1_energy_infidelity} shows the energy and infidelity data after training as function of $h/J$, respectively.
The infidelity with the symmetric ground state suddenly jumps to $F \approx 10^{-1}$ for the symmetry broken states.
The reason why the observables in \Cref{fig:3} were still relatively close to the exact values despite low fidelity is that our minimization objective, the energy expectation value, has the form $E_\theta = -J \cdot C_{zz}(d=1) - h \langle \sigma_x \rangle$.
The Hamiltonian is precisely the sum of $zz$ and $x$ terms and thus the symmetry broken states in fact optimize the sum of both observables.

By setting an initial negative bias with respect to the neurons' activation functions one can steer the variational algorithm to converge to the opposite symmetry broken state where visible neurons are collectively inhibited.
Figure~\ref{fig:si_symbreak} compares the learned state for standard bias initialization at the center of the activation functions and for a shift of $\Delta b = -2\,\text{LSB}$ with the symmetrical ground state distribution, confirming this effect.

\begin{table}[t!]
	\centering
	\caption{Parameter settings for learning the ground state with $N=8$ (256 wave function coefficients) at different $h/J$.} \label{tab:phasetrans_params}
	\begin{tabular}{|c|c|c|c|c|c|c|c|}
		\hline
		$h/J$& 0.1 &  0.5&  0.9&  1.0&  1.25& 5.0& 10.0 \\
		\hline
		$N_\mathrm{sample} [10^5]$ &  2&  2&  4&  2&  2& 2 & 2 \\
		\hline
		$N_h$ & 50 & 30 & 40 & 40 & 30 & 20 & 30\\
		\hline
		$\#$weights & 400 & 240 & 320 & 320 & 240 & 160 & 240\\
		\hline
		$\#$biases & 58 & 38 & 48 & 48 & 38 & 28 & 38\\
		\hline
	\end{tabular}
\end{table}

\section{Details on the choice of network parameters} \label[app]{sec:a_figsi}

Here we specify the network and learning parameters used to produce the data shown in the figures in the main text.

\begin{itemize}
    \item \textbf{Figure 3 (Ising phase transition):}
Each data point used slightly different network topologies and sampling parameters which are summarized in Table~\ref{tab:phasetrans_params}.
Note that these parameters were not optimized and most models are overparameterized with respect to the Hilbert space dimension and likely oversampled.
For $h=0.9$ more samples were required in order to adequately learn both modes of the symmetric ground state.
    \item \textbf{Figure 4 (system-size dependence):}
Learning is performed in a network with $N_h = 40$ hidden units and $N_\mathrm{sample} = 2 \cdot 10^5$ samples are drawn in each iteration for $N=\{3, \ldots, 8 \}$. For $N \in \{9, 10\}$ slightly more hidden units $N_h = 50$ and $N_\mathrm{sample}= 4\cdot 10^5$ samples were used.
    \item \textbf{Figure 5c (weight resolution):}
The uniformly random weights are rounded to subgrids of the full resolution grid $\{-63, -62, \ldots, 0, \ldots, 62,  63\}$ with equidistant steps of sizes $\Delta w \in \{2, 4, 8, 16, 32, 64\}$.
We construct the grid starting at zero and counting up to $64$ in $\Delta w$-steps.
Since the maximum possible weight value is $63$ we decremented the grid edges from $\pm 64$ to $\pm 63$.
Thus, the resulting grids have $128/\Delta w + 1$ possible weight values.
Note that $\Delta w = 1$ represents full resolution with 127 weight values.
\end{itemize}

\section{Analog parameters on BrainScaleS-2}\label[app]{sec:a_bssparams}

Synaptic weights are implemented by two distinct circuits for 6-bit resolved inhibitory ($w_{ij} < 0$) and excitory ($w_{ij} > 0$) interactions.
The biases are regulated by adjusting the leak voltage $V_l$ which is controlled by a 10-bit parameter.
Due to the circuit design we use less than that as can be seen in \cref{fig:1}c where the domain of the activation functions is restricted to around 8-bit.
In general these numbers have to be treated as upper bounds due to the analog nature of the system.
These design choices are not arbitrary, but reflect various trade-offs between chip resources, such as the amount of memory that stores these parameters, their inherent variability and the precision required by applications.

\bibliography{references}{}
\bibliographystyle{apsrev4-1}

\end{document}